\documentclass[aps,prl,twocolumn,reprint,floats,epsfig,showpacs,groupedaddress,superscriptaddress]{revtex4}

\usepackage{bbm}
\usepackage{amssymb}
\usepackage{amsmath}
\usepackage{ifpdf}
\usepackage{color}
\usepackage{graphicx,epsfig}
\usepackage{hyperref}

\newcommand{\Okin}{{\cal O}_{\rm kin}}
\newcommand{\Opot}{{\cal O}_{\rm pot}}

\graphicspath{
{./FIGS/}
}

\begin{document}
\title{Quantum scars from zero modes in an Abelian lattice gauge theory on ladders}

\author{Debasish Banerjee}
\affiliation{Institut f\"ur Physik, Humboldt-Universit\"at zu Berlin, Zum Gro\ss en Windkanal 6, 12489 Berlin, Germany}
\affiliation{Saha Institute of Nuclear Physics, HBNI, 1/AF Bidhannagar, Kolkata 700064, India}
\author{Arnab Sen}
\affiliation{School of Physical Sciences, Indian Association for the Cultivation of Science, Kolkata 700032, India}

\begin{abstract}
  We consider the spectrum of a $U(1)$ quantum link model where gauge fields
  are realized as $S=1/2$ spins and demonstrate a new mechanism
  for generating quantum many-body scars (high-energy eigenstates that violate
  the eigenstate thermalization hypothesis) in a constrained Hilbert space.
  Many-body dynamics with local constraints has attracted much attention
  due to the recent discovery of non-ergodic behavior in quantum simulators
  based on Rydberg atoms. Lattice gauge theories provide natural examples of
  constrained systems since physical states must be gauge-invariant. In our
  case, the Hamiltonian $H=\Okin+\lambda \Opot$, where $\Opot$ ($\Okin$) is
  diagonal (off-diagonal) in the electric flux basis, contains exact
  mid-spectrum zero modes at $\lambda=0$ whose number grows exponentially
  with system size. This massive degeneracy is lifted at any non-zero
  $\lambda$ but some special linear combinations that simultaneously
  diagonalize $\Okin$ and $\Opot$ survive as quantum many-body scars, suggesting
  an ``order-by-disorder'' mechanism in the Hilbert space. We give evidence
  for such scars and show their dynamical consequences on two-leg ladders with
  up to $56$ spins, which may be tested using available proposals of quantum
  simulators. Results on wider ladders point towards their presence in
  two dimensions as well.
 
\end{abstract}
\maketitle

{\it{Introduction}}: Eigenstate thermalization hypothesis (ETH) posits that
individual energy eigenstates of generic many-body systems have ``thermal''
expectation values for local observables with temperature determined by the
energy density of the eigenstate~\cite{ETH1, ETH2, ETH3, ETH4, ETH5}. It also
provides an explanation for the local equilibration of such systems under
their own coherent dynamics~\cite{ETH6, ETH7}. It is equally interesting to
ask when this paradigm fails so that an interacting system
may evade ergodicity. Two
well-known mechanisms are provided by integrability~\cite{VR2016} and
many-body localization~\cite{PH2010,NH2015}. In both cases, an extensive
number of local integrals of motion emerge which prevents the bulk of the
eigenstates from following ETH.

An important question is whether violations of ETH can occur in non-integrable
systems without disorder~\cite{MB2017}. Recently, quench experiments with a
kinematically-constrained chain of $51$ Rydberg atoms~\cite{ryd_exp} exhibited
persistent many-body revivals when initialized in a N\'eel state while, in
contrast, other high-energy initial states thermalized rapidly. Subsequent
theoretical investigations~\cite{pxp1, pxp2} of a minimal model with a
constrained Hilbert space to incorporate strong Rydberg blockade, the PXP
model~\cite{pxp1, pxp2, pxp3, pxp4}, showed this ergodicity-breaking
mechanism is due to the presence of some highly athermal ETH-violating states,
dubbed quantum many-body scars (QMBS), embedded in an otherwise ETH-satisfying
spectrum. A flurry of theoretical research has now shown QMBS to occur in a
variety of other settings~\cite{aklt1,aklt2,aklt3,aklt4,aklt5, qh1,qh2, HM1,
HM2,HM3,HM4, mag1, dyn1,dyn2,dyn3,dyn4,dyn5,dyn6,dyn7, fracton1,fracton2, 
2d_ryd1,2d_ryd2, geometric1, geometric2, flatband} (for a review, see 
Ref.~\onlinecite{scarsreview}).

\begin{figure}[!htb]
  \includegraphics[width=0.36\textwidth]{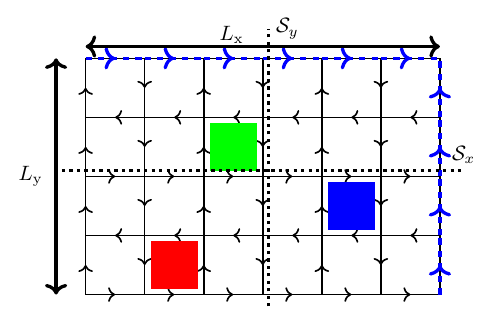} \\
  \includegraphics[scale=0.22]{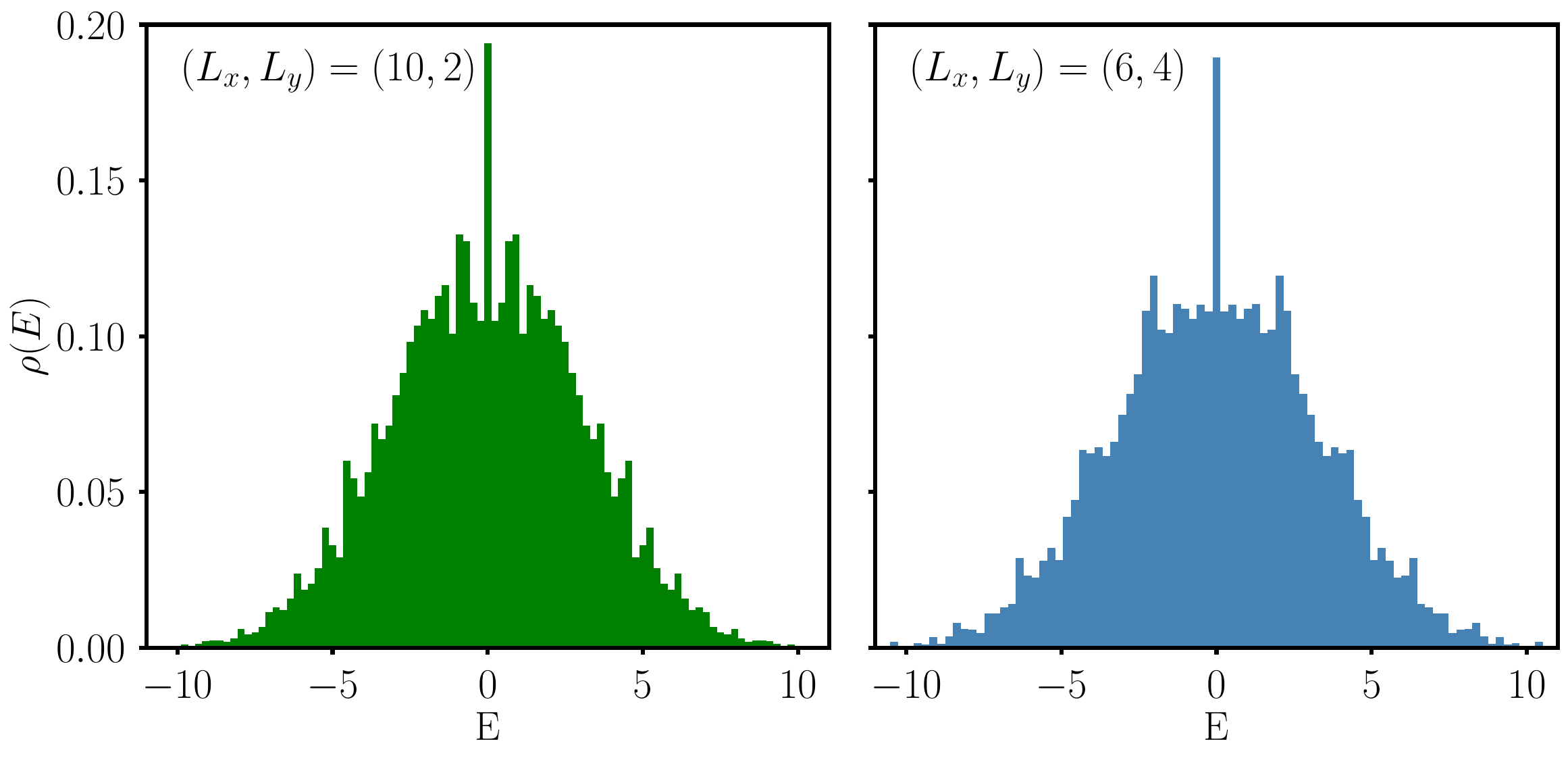}
  \caption{(Top panel) An electric flux configuration for a ladder geometry
    with $(L_x,L_y)=(6,4)$ with periodic boundary conditions in both
    directions. An elementary plaquette with clockwise (anticlockwise)
    circulation of flux is shown in blue (red) while a non-flippable plaquette
    is shown in green. 
    (Bottom panel) The density of states $\rho(E)$ as a function of $E$ for
    $(L_x,L_y) = (10,2)$ (left) and $(6,4)$ (right) at
    coupling $\lambda=0$.}
  \label{fig1}
\end{figure}

Constrained Hilbert spaces arise in Hamiltonian formulations of lattice gauge
theories (LGTs)~\cite{KS1975} since physical (gauge-invariant) states satisfy
an appropriate Gauss law. In fact, the archetypal model for scarring in constrained 
Hilbert spaces, the PXP model, maps exactly to a lattice Schwinger model where 
the gauge fields are coupled with staggered fermions~\cite{PXP_LGT} in one 
dimension. Persistent oscillations starting from the N\'eel state then 
corresponds to a string inversion~\cite{PXP_LGT}.

It is intriguing to ask whether QMBS can appear via a completely different
mechanism in other LGTs. In this Letter, we answer this question in the
affirmative by considering a prototypical LGT but without any
dynamical matter, a $U(1)$ quantum link model (QLM) in spin $S=1/2$
representation~\cite{qlm1,qlm2} on ladder geometries. This $U(1)$ QLM also
arises
as low-energy descriptions of some paradigmatic quantum spin
liquids~\cite{3dqsl, NS2004}, and displays novel crystalline confining
phases~\cite{DB2013}. We show the presence of an exponentially large number
(in system size) of exact mid-spectrum zero modes in a particular limit of
this LGT. Turning on another non-commuting gauge-invariant interaction lifts
this
massive degeneracy but also creates certain special linear combination of
these zero modes that are simultaneous eigenstates of {\it both} the
non-commuting terms in the Hamiltonian. These QMBS are much more
{\it localized} in the Hilbert space compared to the individual zero modes
(which are delocalized)
and hence the mechanism is akin to ``order-by-disorder'', first introduced in
the context of frustrated magnets~\cite{obd1,obd2} (where thermal or quantum
fluctuations lift the exponentially large degeneracy of
classical ground states and makes the system more ordered~\cite{obd3,obd4}),
but in the Hilbert space. The number of such scars depends sensitively
on the ladder geometry with striking differences between ladders of width 
two and four. 

{\it{$U(1)$ QLM}}: We consider the $U(1)$ QLM with the gauge degrees of freedom 
being quantum spins $S = 1/2$ living on the links $\mathbf{r}, \hat{\mu}$ 
connecting two neighboring sites $\mathbf{r}$ and $\mathbf{r}+\hat{\mu}$ (with 
$\hat{\mu} = \hat{i},\hat{j}$) of a ladder of width $L_y$ and length $L_x$ 
(where $L_x$ and $L_y$ are both even), and periodic boundary conditions in both 
directions (Fig.~\ref{fig1}, top panel). A $U(1)$ quantum link, $U_{\mathbf{r},
\hat{\mu}} = S^{+}_{\mathbf{r},\hat{\mu}}$ is a raising operator of the electric 
flux $E_{\mathbf{r},\hat{\mu}} = S^z_{\mathbf{r},\hat{\mu}}$, and the Hamiltonian is
\begin{eqnarray}
  H &=& \Okin+\lambda \Opot \nonumber \\
    &=& -\sum_{\Box}\left( U_\Box + U^\dagger_\Box \right)+\lambda \sum_{\Box}\left( U_\Box + U^\dagger_\Box \right)^2,
  \label{u1qlmH}
\end{eqnarray}
where $U_{\Box} = U_{\mathbf{r},\hat{i}}U_{\mathbf{r}+\hat{i},\hat{j}}U^\dagger_{\mathbf{r}+\hat{j},\hat{i}}U^\dagger_{\mathbf{r},\hat{j}}$ and $\Box$ denotes an 
elementary plaquette. $\Okin$ acts on closed loops of electric flux around 
elementary plaquettes, flipping them from clockwise (anti-clockwise) to 
anti-clockwise (clockwise), while annihilating all other configurations. 
$\Opot$ counts the number of such flippable plaquettes. The Hamiltonian has a 
local $U(1)$ symmetry generated by the Gauss law 
$G_{\mathbf{r}} = \sum_\mu (E_{\mathbf{r},\hat{\mu}}-
E_{\mathbf{r}-\hat{\mu},\hat{\mu}})$. The physical states $|\psi \rangle$ satisfy
$G_{\mathbf{r}}|\psi \rangle =0$ which implies that in- and out-going electric
fluxes add up to zero at each site, thus providing a constrained Hilbert
space (Fig.~\ref{fig1}, top panel).

\begin{figure}[!tbp]
  \includegraphics[width=0.43\textwidth]{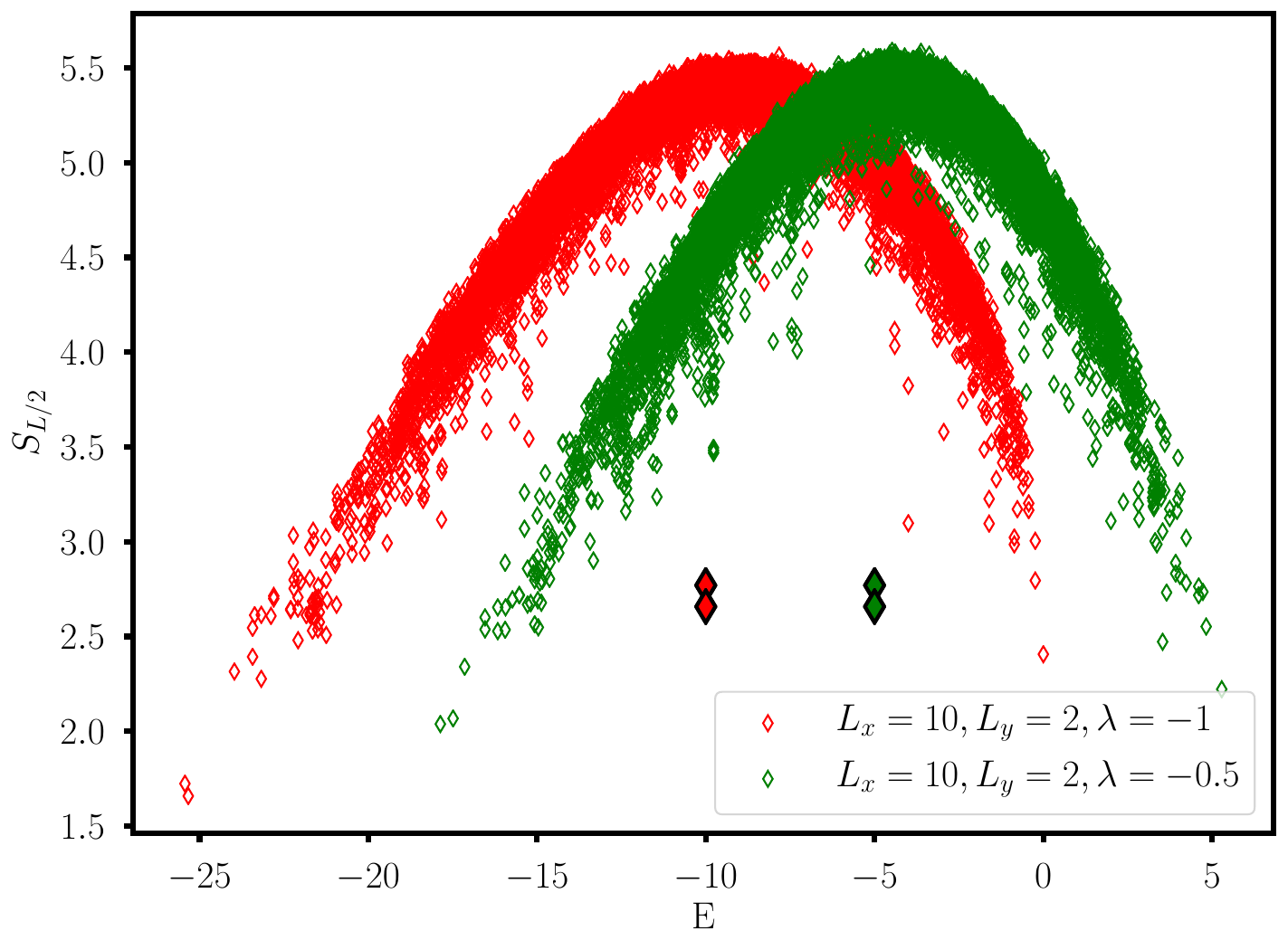}\\
  \includegraphics[width=0.43\textwidth]{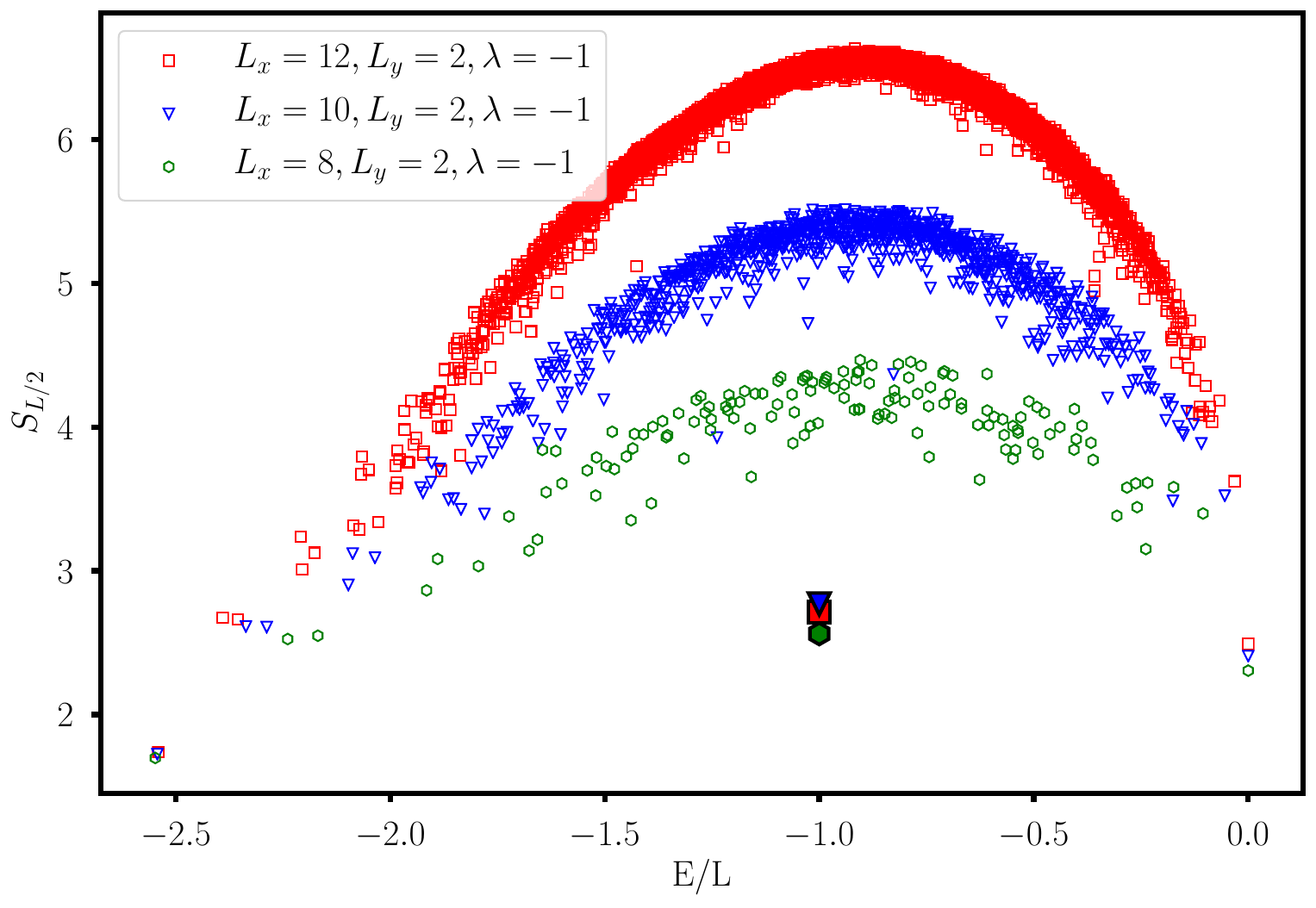}
  \caption{The bipartite entanglement entropy with equal partitions, $S_{L/2}$, 
    shown for all energy eigenstates of the ladder $(L_x,L_y)=(10,2)$ at
    couplings $\lambda=-0.5$ (green) and $\lambda=-1$ 
   (red) (top panel) and for ladders of width $L_y=2$ and $L_x=12,10,8$ 
    (red, blue, green) at $\lambda=-1$ for momentum $(k_x,k_y)=(0,0)$
    (bottom panel).}
\label{fig2}
\end{figure}

The spectrum of $H$ (Eq.~\ref{u1qlmH}) is calculated using large-scale exact
diagonalization (ED). The total electric flux winding around the lattice in a
given periodic direction is a conserved quantity, related to a $U(1)$
center symmetry, and causes the Hilbert space to break up into distinct
topological sectors, characterized by a pair of integer winding numbers
$(W_x,W_y)$. We henceforth restrict ourselves to the largest such sector with
$(W_x,W_y)=(0,0)$. Furthermore, translations by one lattice spacing in
both directions, point-group symmetries like appropriate $180^{\circ}$
rotations and reflections, and charge conjugation (which reverses all
electric fluxes) are discrete symmetries. These symmetries do not mutually
commute and for ED, we use translation symmetry together with the $U(1)$
center symmetry to reach system sizes of up to $(L_x,L_y)=(14,2)$ ($56$ spins)
and $(L_x,L_y)=(8,4)$ ($64$ spins) for ladders of width $L_y=2$ and $L_y=4$,
respectively (see ~\cite{suppmat} for more details).

We calculate the bipartite entanglement entropy 
$S_{L/2} = -\rm{Tr}[\rho_A \ln \rho_A]$ for each energy eigenstate $|\Psi \rangle$
where the reduced density matrix $\rho_A = \rm{Tr}_{B}|\Psi \rangle \langle \Psi|$
is obtained by partitioning the ladder into two equal parts $A$ and
$B$ (see ~\cite{suppmat} for more details), the Shannon entropy
$S_1 = -\sum_\alpha |\psi_\alpha|^2 \ln |\psi_\alpha|^2$ where
$|\Psi \rangle  = \sum_{\alpha=1}^{\mathcal{N}} \psi_\alpha |\alpha\rangle$
when the eigenstate is expressed in a given basis $|\alpha \rangle$ with
$\mathcal{N}$ basis states and the electric flux correlator
$\frac{1}{L_x} \sum_x \langle E_{\hat{j}}(x) E_{\hat{j}}(x+\hat{i}) \rangle$
where $E_{\hat{j}}(x) = \sum_y E_{\mathbf{r},\hat{j}}$. It is sufficient to study
$\lambda \leq 0$ since a unitary transformation relates $H(\lambda)$ and
$H(-\lambda)$~\cite{powell_qdm}. The energy level spacing distribution of the
$U(1)$ QLM with weak disorder (to remove global symmetries) follows the
Gaussian orthogonal ensemble prediction~\cite{rigol_goe} strongly indicating
that the model (Eq.~\ref{u1qlmH}) is non-integrable when
$|\lambda| \lesssim \mathcal{O}(1)$ (see ~\cite{suppmat} for more details).

{\it{Exact zero modes at $\lambda=0$}}: At $\lambda=0$, the Hamiltonian
anti-commutes with the operator 
$\mathcal{C} = \prod_{\mathbf{r},\hat{\mu}} E_{\mathbf{r},\hat{\mu}}$ where only the
horizontal (vertical) links on even $x$ ($y$) contribute to the product and
thus each elementary plaquette contains precisely one such link. This implies
that any eigenstate with energy $E \neq 0$ has a partner at
$-E$ $(\mathcal{C}|E\rangle = |-E\rangle)$. For any $\lambda$, $H$ also
commutes with space reflections about the axes $\mathcal{S}_x, \mathcal{S}_y$
(Fig.~\ref{fig1}, top panel). This point-group symmetry 
commutes with $\mathcal{C}$. Remarkably, any Hamiltonian with these properties
has exact zero-energy eigenstates whose number scales exponentially in the
system size due to an index theorem~\cite{indexth}. Since the spectrum is
symmetric around $E=0$, these protected zero modes are mid-spectrum states of
the Hamiltonian and not low-energy states bound to topological
defects or arising from supersymmetry for which there are
well-known examples~\cite{zm1,zm2,zm3,zm4,zm5,zm6}.
The presence of a large number of mid-spectrum zero modes is clear from the
density of states $\rho(E)$ as a function of energy $E$, which shows a sharp
spike at $E=0$ from the (momentum unresolved) data generated for both
$(L_x,L_y)=(10,2)$ and $(6,4)$ (Fig.~\ref{fig1}, bottom panel). Despite there
being manifestly no level repulsion between them, the behavior of the Shannon
entropy, $S_1$, for the zero modes indicates that these are not
anomalous when compared to other neighboring eigenstates~\cite{indexth}
(see ~\cite{suppmat} for more details). Thus, generic
  linear combinations of such zero modes are expected to be delocalized in
  the Hilbert space and have volume-law entanglement.

{\it{QMBS at $\lambda \neq 0$}}: At a non-zero $\lambda$, $\mathcal{C}$ no
longer anti-commutes with the Hamiltonian and hence the manifold of zero modes
is not protected and mix with the non-zero modes to form new high-energy
eigenstates. The only possible exception are special
linear combinations of the zero
modes which are also eigenstates of $\Opot$, and hence of $H$.
Remarkably, these linear combinations also show localization in the Hilbert 
space and anomalously low entanglement. Such
eigenstates clearly violate ETH since they remain unchanged as the coupling
$\lambda$ is varied in spite of the energy level spacing in their neighborhood
being exponentially small in $L_xL_y$. 

{\it{Ladders with $L_y=2$}}: Using ED for linear dimension $8 \leq L_x \leq 14$, 
we show the presence of $4$ QMBS, one each at momenta $(k_x,k_y)=(0,0),(\pi,\pi),
(\pi,0)$ and $(0,\pi)$ where each of these scars is a simultaneous eigenfunction 
of $\Opot$ (with eigenvalue $N_p/2$, $N_p$ being the total number of plaquettes) 
and $\Okin$ (with eigenvalue $0$).

A defining property of QMBS is that they have a much lower entanglement entropy 
compared to their neighboring energy eigenstates and thus show up as entropy-outliers 
in $S_{L/2}$. This is demonstrated in the momentum-unresolved data for $S_{L/2}$ at 
$L_x=10$ where the $4$ QMBS are clearly visible as outliers (each with a double 
degeneracy) at both $\lambda=-0.5$ and $\lambda=-1$ (Fig.~\ref{fig2}, top panel). 
All $4$ scars have the energy $E=\lambda N_p/2$. Restricting to $(k_x,k_y)=(0,0)$ 
and comparing the data of $S_{L/2}$ for $L_x=12, 10, 8$ at $\lambda=-1$ 
(Fig.~\ref{fig2}, bottom panel) shows that while the other mid-spectrum eigenstates 
seem to follow a volume law scaling for $S_{L/2}$ as expected from ETH, the scar
state has a much lower $S_{L/2}$ that scales anomalously.

\begin{figure}
 \includegraphics[width=0.42\textwidth]{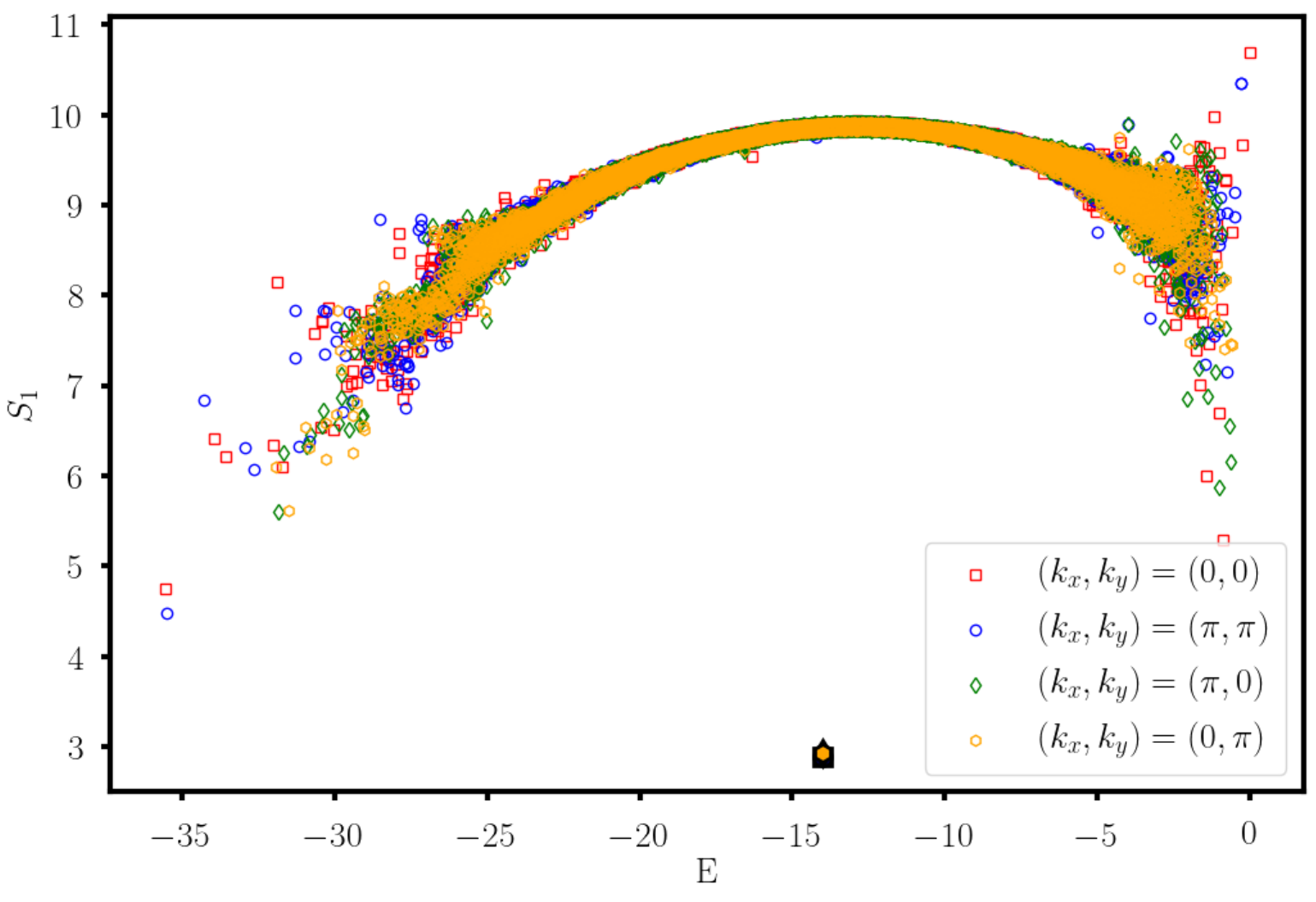}\\
 \includegraphics[width=0.47\textwidth]{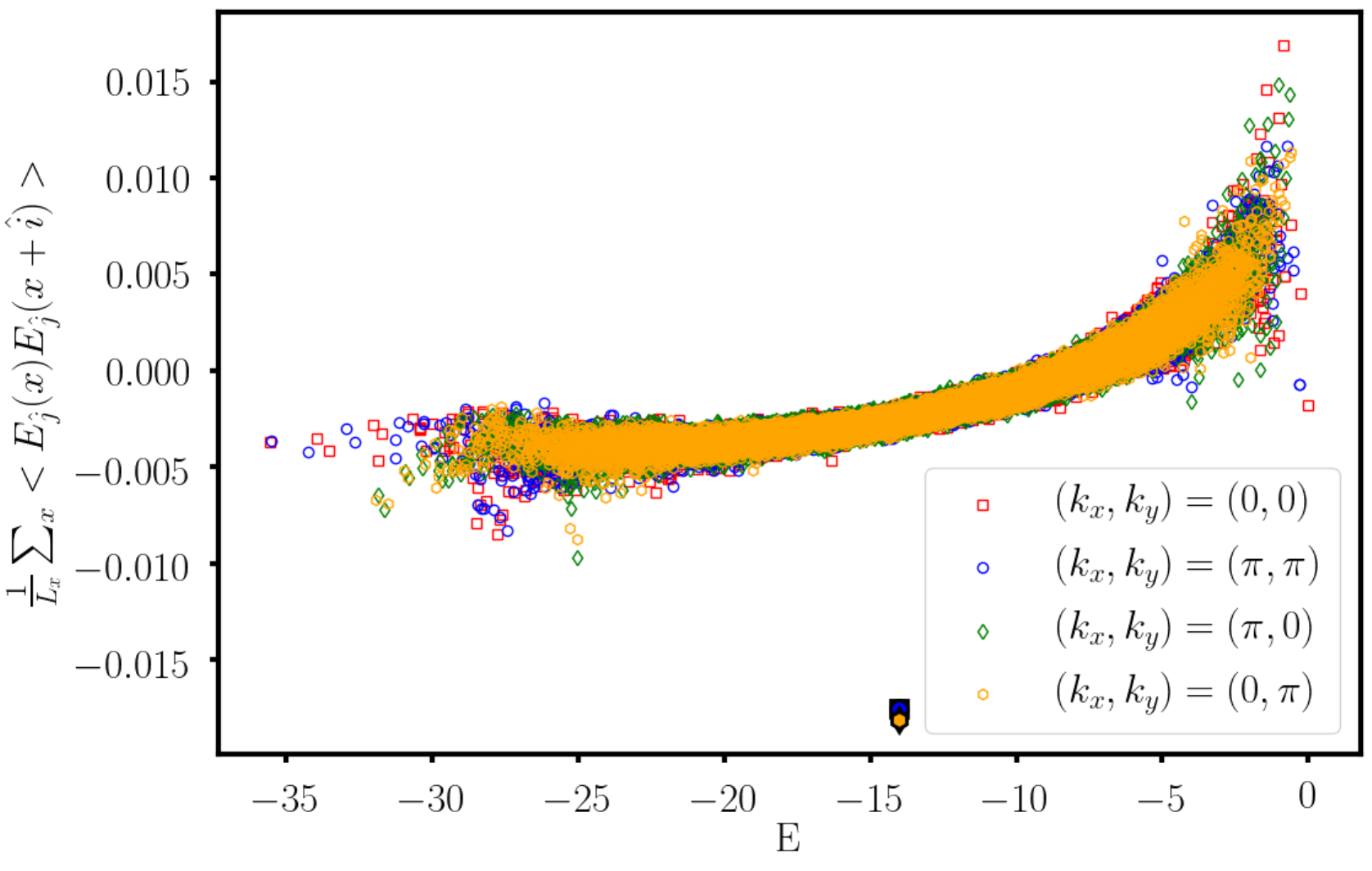}
 \caption{The Shannon entropy, $S_1$, (top panel) and the electric flux correlator 
   (bottom panel) for all energy eigenstates of $(L_x,L_y) = (14,2)$ at
   momenta $(k_x,k_y)= (0,0), (\pi,\pi), (\pi,0), 
 (0,\pi)$ (shown in red, blue, green, orange) with coupling $\lambda=-1$.}
\label{fig3}
\end{figure}

Further evidence for scarring is provided by the behavior of the Shannon entropy 
$S_1$ and the electric flux correlator at $L_x=14$. In Fig.~\ref{fig3} (top panel), 
we show the $S_1$ for the momentum-resolved eigenstates at momenta $(k_x,k_y)=(0,0), 
(\pi,\pi), (\pi,0)$ and $(0,\pi)$ from which it is clear that while the neighboring 
eigenstates with similar energies are delocalized amongst the basis states as 
expected of high-energy states, the $4$ QMBS have a much lower $S_1$ indicating 
their localization in the Hilbert space. The electric flux correlator for the QMBS 
have markedly different values from the neighboring energy eigenstates (Fig.~\ref{fig3}, bottom panel).


That special linear combinations of the exact zero modes are responsible for the 
creation of these QMBS can be verified by diagonalizing $\Opot$ only in the zero mode 
subspace. While most of the resulting eigenvalues are non-integers, which indicate 
that these states must mix with non-zero modes at finite $\lambda$ (since $\Opot$ 
is a counting operator), there is one eigenvector at each of the momenta 
$(k_x,k_y)=(0,0),(\pi,\pi),(\pi,0)$ and $(0,\pi)$ with an integer eigenvalue of 
$\Opot$ that equals $N_p/2$. We have also checked that the corresponding eigenvectors 
have exactly the same wavefunctions as the QMBS generated at finite $\lambda$ 
(see ~\cite{suppmat} for more details).

Expressing any of these QMBS in terms of the zero modes shows that they appear 
to be a pseudo-random superposition of all the zero modes (see ~\cite{suppmat} for 
more details). However, these states have a much smaller $S_1$ compared to any 
of the individual zero modes (see ~\cite{suppmat} for more details) showing that 
they are much more localized in the Hilbert space. The $\Opot$ term thus induces 
a subtle order-by-disorder mechanism in the exponentially large subspace of the 
zero modes to pick out a few special linear combinations and generate the QMBS at 
$\lambda \neq 0$.

\begin{figure}
 \centering
 \includegraphics[width=0.43\textwidth]{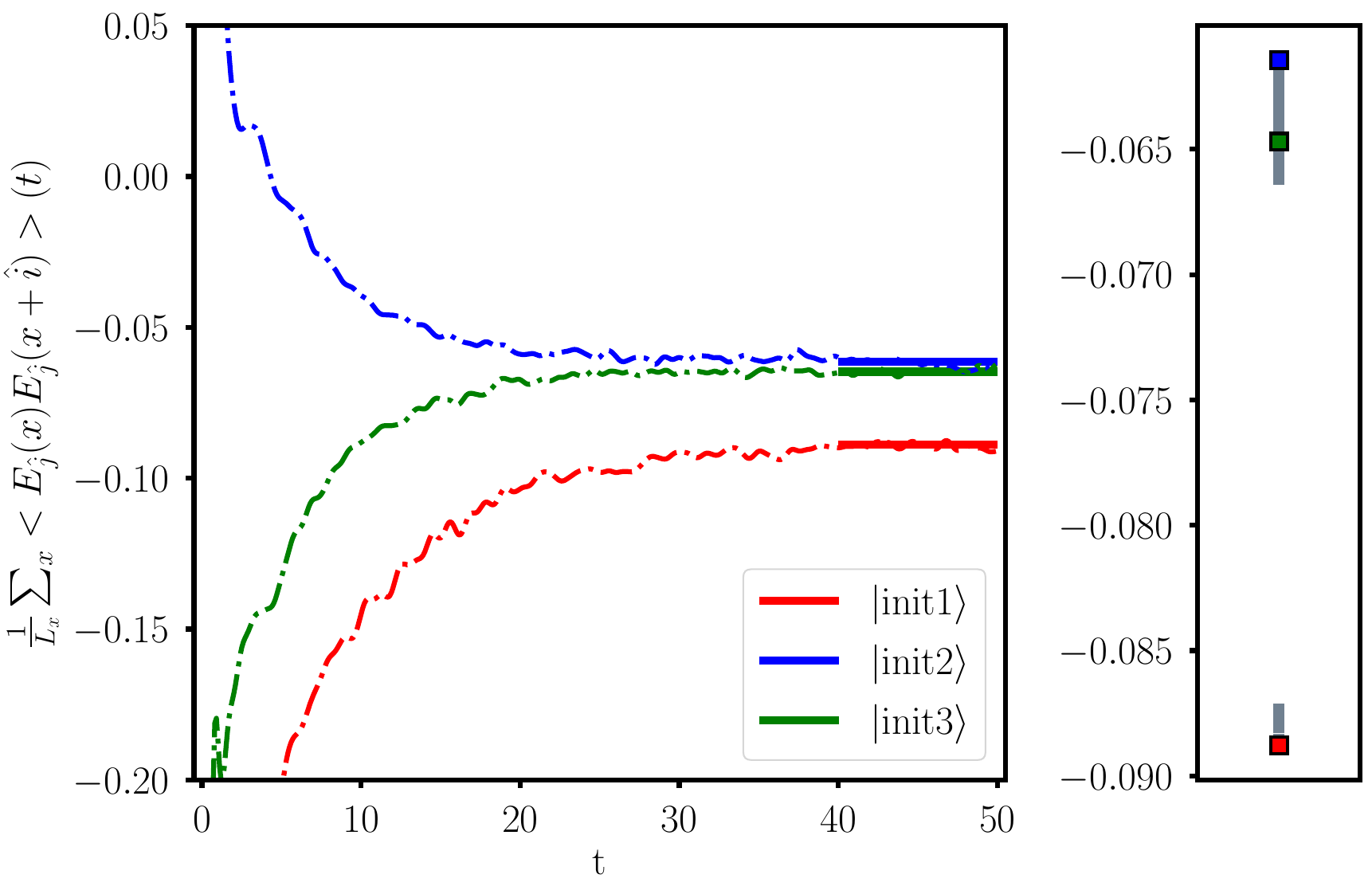}\\
 \includegraphics[width=0.43\textwidth]{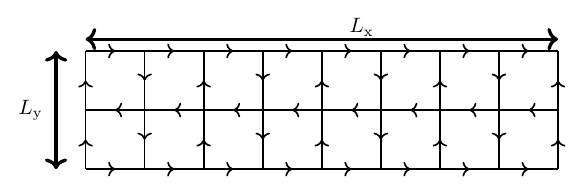}
 \caption{(Top left panel) Unitary dynamics for $(L_x,L_y)=(14,2)$ with three
   simple initial states $|\mathrm{init}1,2,3\rangle$ ($|\mathrm{init1} \rangle$ 
   has a finite overlap with the QMBS at $(k_x,k_y)=(0,0)$, while 
   $|\mathrm{init2,3}\rangle$ has zero overlap) showing the behavior of the 
   electric flux correlator at $\lambda=-1$ as a function of time. (Top right panel)
   Late-time value of the electric flux correlator for the $6433$ initial states 
   with $(k_x,k_y)=(0,0)$ and average energy $\lambda N_p/2$ at $(L_x,L_y)=(14,2)$.
  (Bottom panel) A simple reference state that generates momentum eigenstates 
   with finite overlap with QMBS for $L_x=8,10,12,14$.}
 \label{fig4}
 \end{figure}

These QMBS also leave an imprint on the unitary dynamics starting from simple 
high-energy initial states. Consider the class of all $(k_x,k_y)=(0,0)$ initial 
states generated from any reference electric flux state with $N_p/2$ flippable 
plaquettes. For $(L_x,L_y)=(14,2)$, there are $6433$ such initial states with 
identical average energy of $\lambda N_p/2$. Out of these, only $18$ initial 
states have a non-zero overlap with the QMBS at $(k_x,k_y)=(0,0)$. In 
Fig.~\ref{fig4} (top left panel), we consider $\lambda=-1$ and 
$|\mathrm{init1} \rangle$, an initial state with a finite overlap with the 
QMBS while $|\mathrm{init2}\rangle$, $|\mathrm{init3}\rangle$ have zero overlap. 
While the latter two initial states, even with very different starting values 
of the electric flux correlator, converge to very similar values at late times 
indicating ETH-guided thermalization, $|\mathrm{init1} \rangle$ instead converges 
to a very different (subthermal) value showing that the state retains memory 
of its overlap with the scar at late times. The late-time value of the electric 
flux correlator for all these $6433$ initial states (Fig.~\ref{fig4} (top 
right panel)) clearly shows that the $18$ initial states with non-zero overlap 
with the QMBS converge to markedly different values compared to the rest.
Finally, the data for $L_x=8,10,12,14$ shows that a simple initial reference 
state with a $2 \times 2$ unit cell (Fig.~\ref{fig4}, bottom panel) always 
generates $2$ momentum eigenstates with finite overlap with $2$ of the QMBS 
(at momenta $(\pi,0), (0,\pi)$ for $L_x=14,10$ and $(0,0), (\pi,\pi)$ for $L_x=12,8$).

 \begin{figure}[!tbp]
 \centering
 \includegraphics[width=0.45\textwidth]{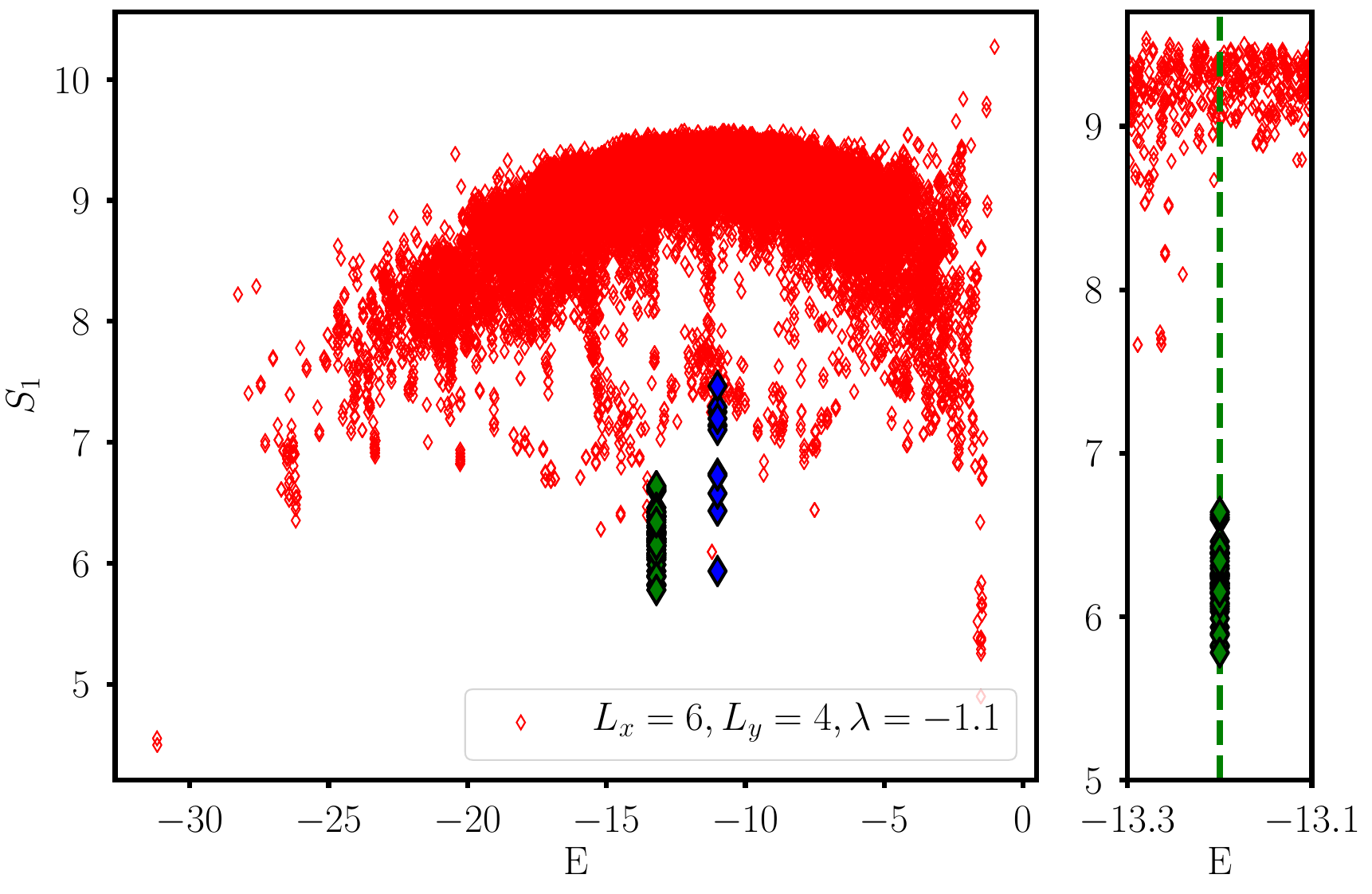}
 \caption{(Left panel) The Shannon entropy, $S_1$, for all the eigenstates for 
 $(L_x,L_y)=(6,4)$. The $46$ ($12$) eigenstates with eigenvalues $(N_p/2,0)$ 
 ($(5N_p/12,0)$) for $(\Opot,\Okin)$ are shown in green (blue) for $\lambda=-1.1$. 
 (Right panel) The same data shown in the vicinity of $E=\lambda N_p/2$ (dotted line).}
 \label{fig5}
 \end{figure}

 {\it{Ladders with $L_y=4$}}: We consider $L_y=4$ ladders to approach closer to the 
 two-dimensional limit.
 The momentum-unresolved data for all the $32810$ eigenstates at $L_x=6$ strikingly 
shows that the number of QMBS is sensitive to the width of the ladder with there being 
$46$ such anomalous eigenstates with the eigenvalue $(N_p/2,0)$ and $12$ with eigenvalue 
$(5N_p/12,0)$ for $(\Opot,\Okin)$, respectively (Fig.~\ref{fig5} (left panel)) (see  \cite{suppmat} for data at $L_x=8$ for
$(k_x,k_y)=(0,0), (\pi,\pi), (\pi,0)$ and $(0,\pi)$ and evidence of QMBS therein). As shown 
in Fig.~\ref{fig5} (right panel), these $46$ eigenstates are degenerate with
$E=\lambda N_p/2$ and have a lower value of $S_1$ compared to that of other neighboring
eigenstates which have $E \neq \lambda N_p/2$. Diagonalizing $\Opot$ in the zero mode 
subspace also yields exactly $46$ ($12$) eigenvectors with the right integer eigenvalues 
$N_p/2$ ($5N_p/12$) with the other eigenvalues being non-integers.

{\it{Conclusions and outlook}}: We have considered a $U(1)$ QLM on finite ladders 
and demonstrated a new mechanism for the formation of QMBS. For the limit when the 
Hamiltonian only contains off-diagonal terms in the electric flux basis, there is 
an exponentially large (in system size) manifold of exact zero modes due to an 
index theorem. This massive degeneracy is lifted by applying another gauge-invariant 
interaction, diagonal in the electric flux basis, but some special linear combinations
of the zero modes that simultaneously diagonalize both the non-commuting terms 
in the Hamiltonian survive and form QMBS. These also leave an imprint on the 
coherent dynamics of this LGT and leads to the absence of thermalization from 
a class of simple initial states, in particular one that can be generated from a 
reference state with a $2 \times 2$ unit cell. This effect can, in principle, be 
verified on quantum simulators based on superconducting qubits or Rydberg arrays
using existing proposals~\cite{expLGT1, expLGT2}. One possible
way is to use the duality transformation to rewrite the Hamiltonian as a quantum
Ising spin model. The disallowed plaquette flips can then be forbidden via Rydberg 
blockade to realize a gauge invariant interaction~\cite{expLGT2}.

 Several open issues arise from our work. Whether this mechanism survives in higher 
dimensions is an obvious question given our results on  wider ladders. Do QMBS arise in 
non-Abelian QLMs is another interesting direction to explore. An analytic understanding 
of the algebraic properties of the zero modes and these special linear combinations is 
highly desirable to address whether such scarring survives in the thermodynamic limit.
Finally, adding further interactions to models with an exponentially large manifold of 
mid-spectrum zero modes may provide yet another route to QMBS.

{\it{Acknowledgements:}}
  We thank Diptiman Sen, Krishnendu Sengupta and Uwe-Jens Wiese for useful discussions 
  and acknowledge computational resources of DESY on the PAX cluster. We also
  thank Saptarshi Biswas (IISER Kolkata) for spotting a numerical error in an earlier
  version.
D.B. acknowledges support by the German Research Foundation (DFG), Grant ID BA 5847/2-1. 
The work of A.S. is partly supported through the Max Planck Partner Group between the 
Indian Association for the Cultivation of Science (Kolkata) and the Max Planck Institute 
for the Physics of Complex Systems (Dresden).

\section{Supplementary Material for ``Quantum scars from zero modes in an Abelian lattice gauge theory on ladders''}

\section{Aspects of exact diagonalization}
 
 \begin{center}
  \begin{tabular}{|c|c|c|c|}
	  \hline
	  $(L_x, L_y)$ & Gauss law & $(W_x,W_y)=(0,0)$ & $(k_x,k_y)=(0,0)$ \\
	  \hline
          (8,  2)  &      7074 &      2214     &       142    \\
	  (10, 2)  &     61098 &     17906     &       902    \\
	  (12, 2)  &    539634 &    147578     &      6166    \\
          (14, 2)  &   4815738 &   1232454     &    44046     \\
	  (16, 2)  &  43177794 &  10393254     &   324862     \\
          (4,  4)  &      2970 &       990     &       70     \\
          (6,  4)  &     98466 &     32810     &     1384     \\
          (8,  4)  &   3500970 &   1159166     &    36360     \\
          (6,  6)  &  16448400 &   5482716     &   152416     \\
	  \hline
  \end{tabular}
\end{center}
 The unconstrained Hamiltonian with $S=1/2$ links on a $(L_x,L_y)$ ladder contains
 $2^{2L_x L_y}$ configurations. Since gauge-invariant states have the added constraint 
 that in- and out-going electric fluxes add up to zero at each site, this dramatically 
 decreases the number of allowed states in the Hilbert space. Furthermore, restricting 
 to the largest topological sector with $(W_x,W_y)=(0,0)$ reduces the number of allowed 
 configurations even further. Lastly, using the additional global symmetries of 
 translations in both directions, allows access to bigger system sizes as shown in the 
 table for the largest momentum block of $(k_x,k_y)=(0,0)$ for various $(L_x,L_y)$.
 We are able to obtain the full spectrum of Hamiltonians with upto $\sim 75000$
 states, while real-time dynamics is possible for Hamiltonians with upto 
 $\sim 50000$ states.

\section{Calculation of Bipartite Entanglement Entropy}

\begin{figure}[!tbh]
 \begin{center}
 \includegraphics[scale=0.8]{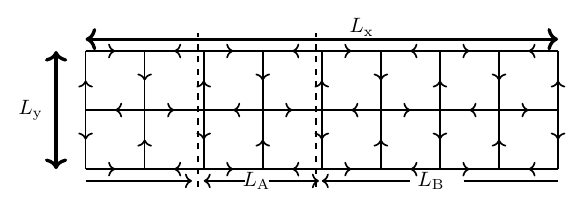}
 \caption{A ladder of dimension $(L_x,L_y)$ is divided into two subsystems of dimensions 
   $L_A$ and $L_B$ along its length for calculating the bipartite entanglement entropy.}
 \label{fig:cart1}
 \end{center}
\end{figure}
We outline computation for the bipartite entanglement entropy in the ladder geometry $(L_x,L_y)$. Let $n_{\rm D}$ 
denote the total of gauge invariant basis states in a given sector $(W_x,W_y)$, where we 
represent the basis as:
\begin{equation}
  \left\{ | e \rangle \right\} = \left\{ | e_k \rangle; k = 1, \cdots, n_{\rm D} \right\},
\end{equation}
and a wavefunction in this sector can be written as a linear superposition
\begin{equation}
  | \psi \rangle = \sum_{k=1}^{n_{\rm D}} c_k | e_k \rangle.
\end{equation}
To compute the bipartite entanglement entropy $S_{\rm A}$ for the partition of the system 
into two subsystems, as sketched in Fig \ref{fig:cart1}, we first denote the basis states 
spanning the two partitions by $\left\{ | e^{(A)} \rangle \right\}$ and  
$\left\{ | e^{(B)} \rangle \right\}$. Note that $\left\{ | e^{(A,B)} \rangle \right\}$ are 
gauge invariant at all points in the bulk regions A and B respectively, but represent surface 
charges at the boundaries.

The calculation of $\left\{ | e^{(A,B)} \rangle \right\}$ is straightforward: for each $k$ in
$\left\{ | e_k \rangle \right\}$, we split the links along the vertical dotted line shown in 
the figure and sort them into $\left\{ | e^{(A)} \rangle \right\}$ and 
$\left\{ | e^{(B)} \rangle \right\}$ respectively, depending on whether they fall into the 
region A or B. Then, we remove the duplicates and obtain:
\begin{equation}
\begin{split}
\left\{ | e^{(A)} \rangle \right\} & = \left\{ | e^{(A)}_{i_A}\rangle, i_A = 1, \cdots, D_A   \right\} \\
\left\{ | e^{(B)} \rangle \right\} & = \left\{ | e^{(B)}_{i_B}\rangle, i_B = 1, \cdots, D_B   \right\}
\end{split}
\end{equation}

Now, for a general energy eigenstate $| \psi \rangle $,
\begin{equation}
\begin{split}
    | \psi \rangle & = \sum_{k}^{n_{\rm D}} c_k | e_{k} \rangle \\
    & = \sum_{i_A}^{D_A} \sum_{i_B}^{D_B} \chi_{i_A, i_B} | e^{(A)}_{i_A} \rangle \otimes | e^{(B)}_{i_B} \rangle \\
    & = \sum_{\ell}^{n_\chi} \chi_\ell | \tilde{e}^{(A)}_{\ell} \rangle \otimes | \tilde{e}^{(B)}_{\ell} \rangle
\end{split}
\end{equation}
First, the individual basis vectors from the two subsystems are "patched" with each other: for 
example the $i_A$ from ${ | e^{(A)}\rangle}$ is patched with $i_B$ from ${ | e^{(B)}\rangle}$ 
to form the corresponding matrix element $\chi (i_A, i_B)$. Clearly, this makes $\chi$ a rectangular 
matrix of dimensions $D_A \times D_B$. To go from the second to the third step, one does Schmidt 
decomposition, which yield the real and non-negative Schmidt values 
$\chi_\ell, \ell = 1, \cdots, n_\chi $, where $n_\chi = {\rm min} ( D_A, D_B )$.

The (von-Neumann) entanglement entropy for this sub-partition of the state $| \psi \rangle$ when 
$L_A=L_B$ then given by
\begin{equation}
  S_{L/2} = -\sum_{i = 1}^{n_\chi} | \chi_i |^2 {\rm ln} \left( | \chi_i |^2  \right)
\end{equation}
and is shown in the main text. 

\section{Level Spacing distribution for weakly disordered $U(1)$  QLM}
\begin{figure}
\includegraphics[width=0.43\textwidth]{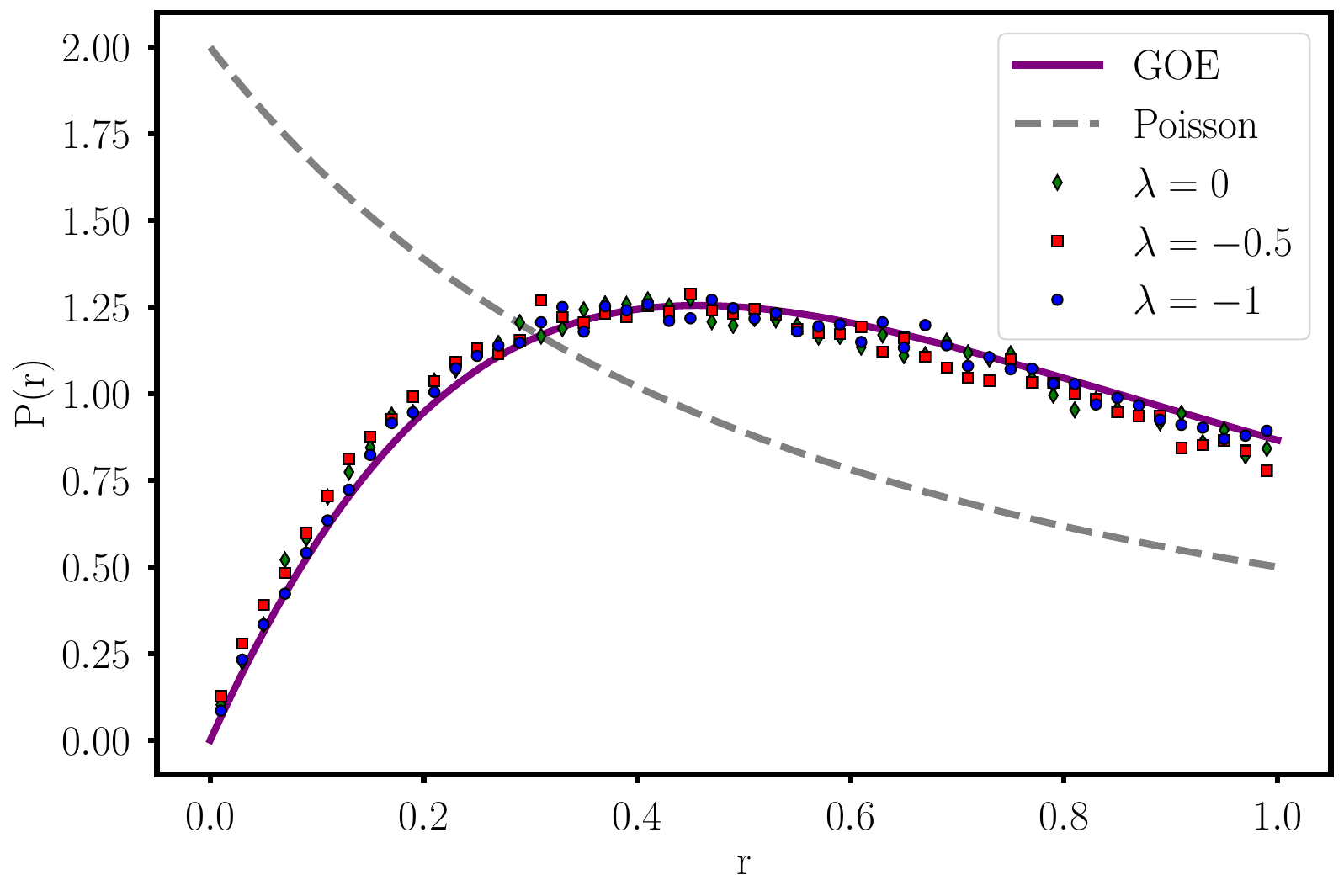}
\caption{The probability distribution, $P(r)$, obtained from the ratio of consecutive energy 
  gaps for the weakly disordered $U(1)$ QLM on a ladder of dimension $(L_x,L_y)=(12,2)$ for 
  $\lambda=0,-0.5,-1.0$ matches well with the prediction from the Gaussian orthogonal ensemble 
  (thick curve), and not the Poisson statistics expected for integrable systems (dotted curve).}
\label{levelspacing}
\end{figure}
To show that the $U(1)$ QLM on ladders is non-integrable in the topological sector $(W_x,W_y)=(0,0)$, 
we take a ladder of dimension $(L_x,L_y)=(12,2)$ and weakly disorder the $\lambda$ term in 
Eq. 1 of main text to $\lambda_i = \lambda (1+\alpha r_i)$ ($\lambda_i = \alpha r_i)$ for non-zero 
(zero) $\lambda$ where $\alpha=0.1$ and $r_i$ is a random number chosen with uniform probability 
between $[-1/2,1/2]$ on the $i$th elementary plaquette. The small $\alpha$ ensures that the global
symmetries of translations and point-group symmetries of reflections and $180^{\circ}$ rotations are 
lifted. We then resolve the energy eigenstates in the only remaining (internal) symmetry of charge
conjugation and focus on the block with eigenvalue of $+1$ (the other block has eigenvalue $-1$) 
which gives $73789$ energy eigenvalues (denoted by $E_n$). We then obtain the probability distribution
$P(r)$, with $r$ being the ratio of two consecutive energy gaps,
\begin{eqnarray}
  r&=& \frac{\mathrm{min}(s_n,s_{n+1})}{\mathrm{max}(s_n,s_{n+1})} \in [0,1] \nonumber \\
  s_n&=&E_{n+1}-E_n
  \label{gaps}
  \end{eqnarray}
For a system satisfying ETH, this distribution is expected to converge to the Gaussian orthogonal 
ensemble, where
\begin{eqnarray}
  P_{GOE}(r) = \frac{27}{4} \frac{r+r^2}{(1+r+r^2)^{5/2}}.
  \label{GOE}
\end{eqnarray}
This indeed seems to be the case for the weakly disordered $U(1)$ QLM for $\lambda=0,-0.5,-1.0$ 
(Fig.~\ref{levelspacing}). On the other hand, the level statistics for an integrable system should follow Poisson statistics with $P(r)=2/(1+r)^2$ (also shown
for reference in Fig.~\ref{levelspacing}).

\section{Shannon entropy of energy eigenstates at $\lambda=0$}
\begin{figure}
\includegraphics[width=\hsize]{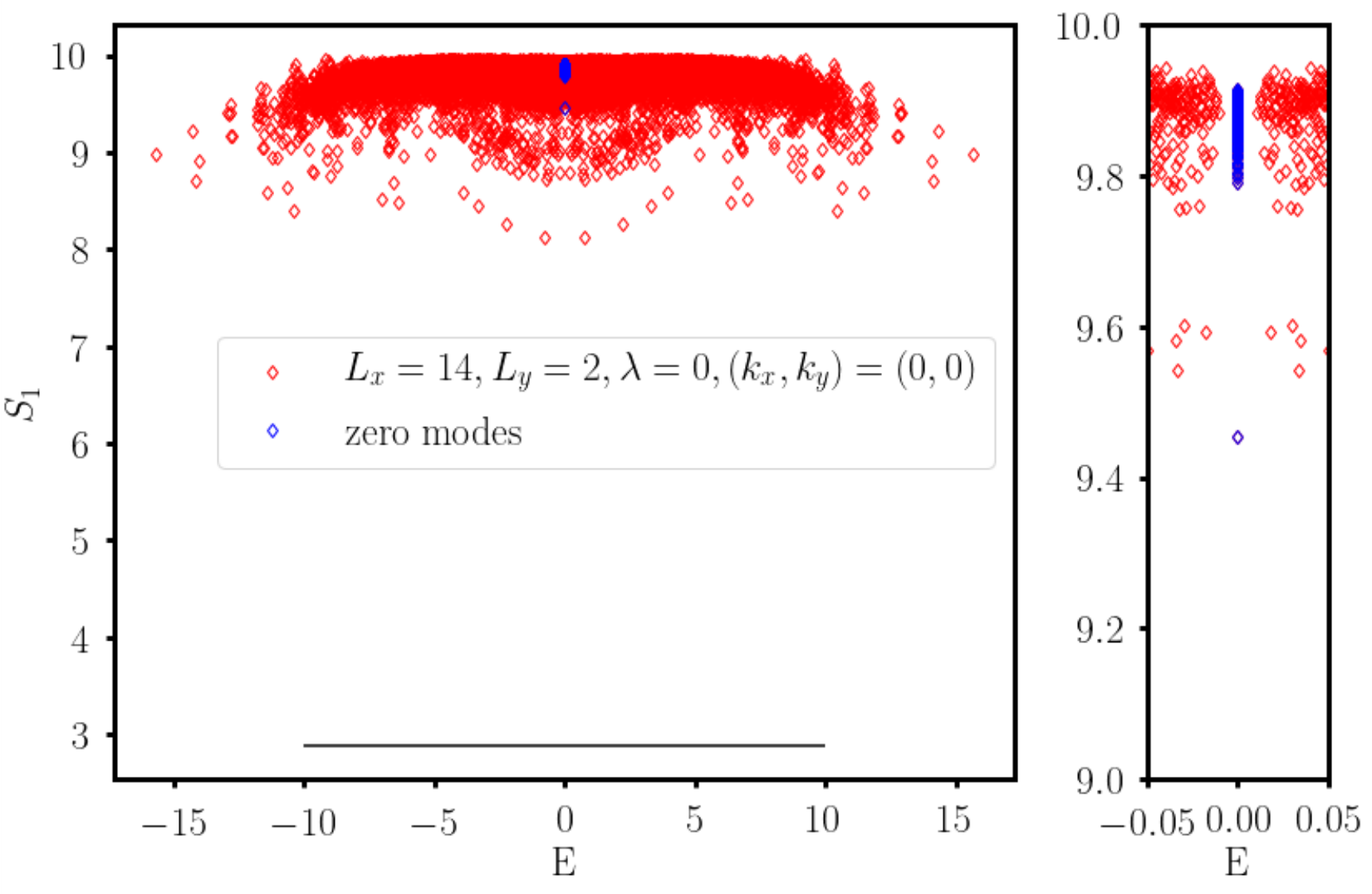}
\caption{(Left panel) The behavior of Shannon entropy, $S_1$, for the energy eigenstates with 
 momentum $(k_x,k_y)=(0,0)$ shown at coupling $\lambda=0$ for ladder dimension of $(L_x,L_y)=(14,2)$.
 The horizontal line equals the value of $S_1$ for the QMBS at finite $\lambda$. (Right panel) The 
 same data of $S_1$ shown in the vicinity of $E=0$. The zero modes are indicated in blue for both 
 panels.}
\label{zeromodes}
\end{figure}

The Shannon entropy, $S_1$, of the energy eigenstates at $\lambda=0$ with momentum $(k_x,k_y)=(0,0)$ 
is shown in Fig.~\ref{zeromodes} (left panel) for a ladder of dimension $(L_x,L_y)=(14,2)$. The $S_1$ 
for the QMBS at $\lambda \neq 0$ is also shown as a horizontal line from which it is clear that 
the QMBS is much more localized in the Hilbert space. Fig.~\ref{zeromodes} (right panel) shows the 
same data for $S_1$ at $\lambda=0$ in the vicinity of $E=0$ with the zero modes indicated in blue, which shows that the zero modes are as delocalized in the Hilbert space as the non-zero 
modes in the neighborhood and do not have an anomalously low $S_1$ like the QMBS (that emerges at 
finite $\lambda$).

\section{QMBS for $(k_x,k_y)=(0,0),(\pi,\pi),(\pi,0), (0,\pi)$ for ladders of width $(L_x,L_y)=(8,4)$}

 \begin{figure}
\includegraphics[width=0.48\textwidth]{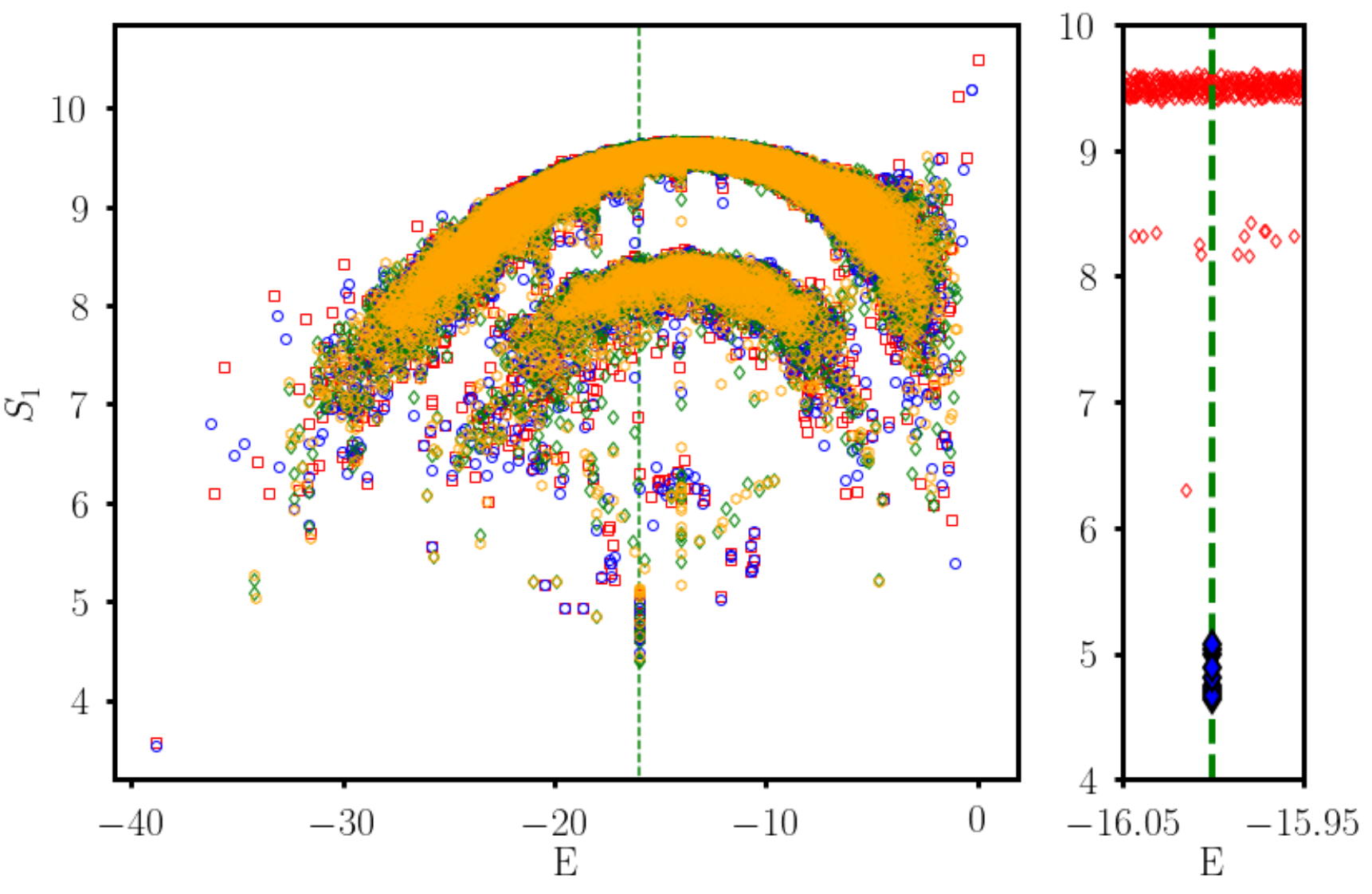}\\
  \includegraphics[width=0.43\textwidth]{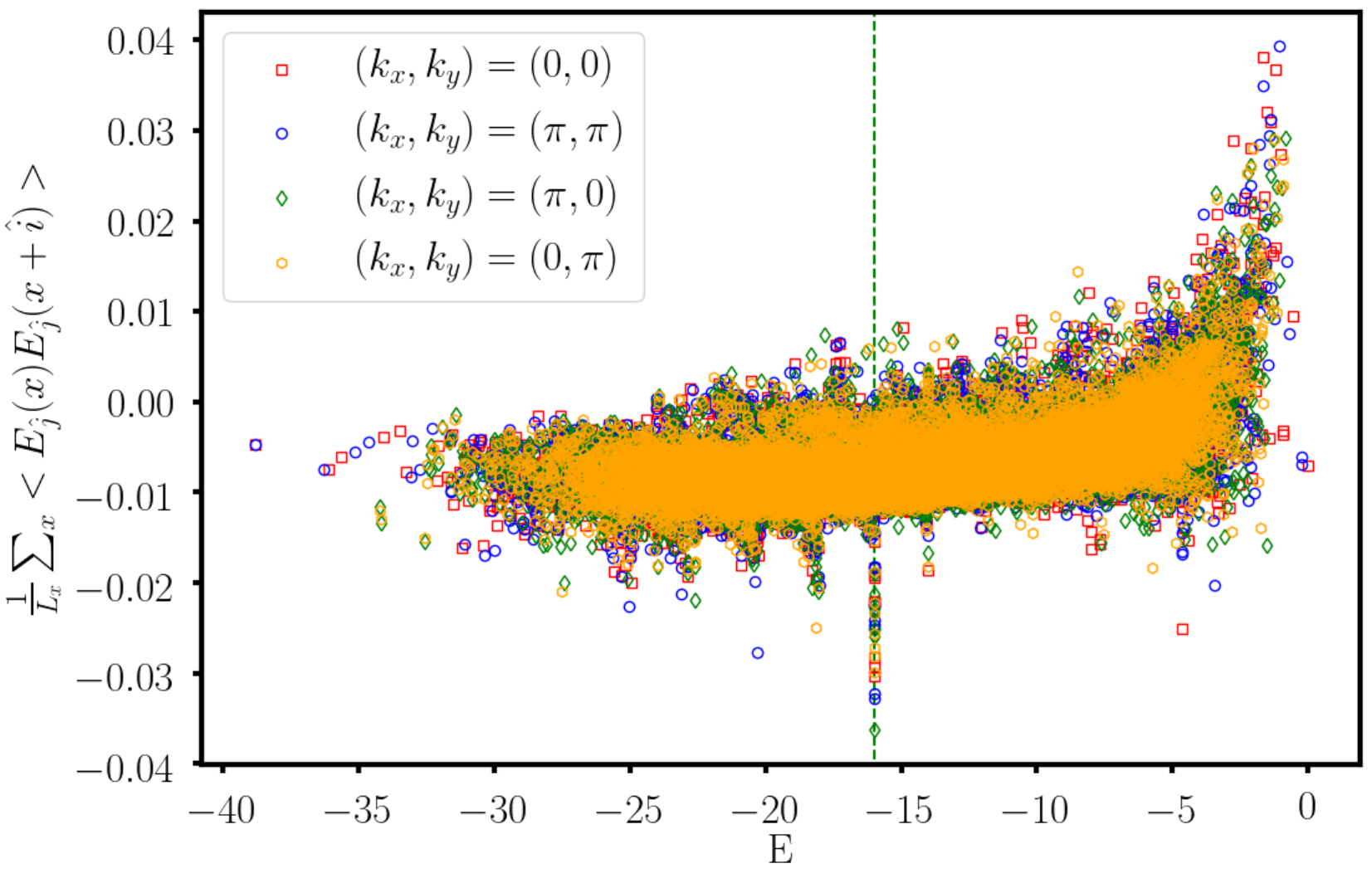}
  \caption{The Shannon entropy, $S_1$, (top left panel) and the electric flux correlator (bottom panel) 
for all energy eigenstates of $(L_x,L_y) = (8,4)$ at $(W_x,W_y)=(0,0)$, and for the momenta 
$(k_x,k_y)= (0,0),(\pi,\pi),(\pi,0),(0,\pi)$ (shown in red, blue, green, orange) with coupling 
$\lambda=-1$. The top right panel shows the same data for $S_1$ for $(k_x,k_y)=(0,0)$ in the vicinity of $\lambda N_p/2$. }
  \label{wideladder}
 \end{figure}

 \begin{figure}
   \includegraphics[width=0.43\textwidth]{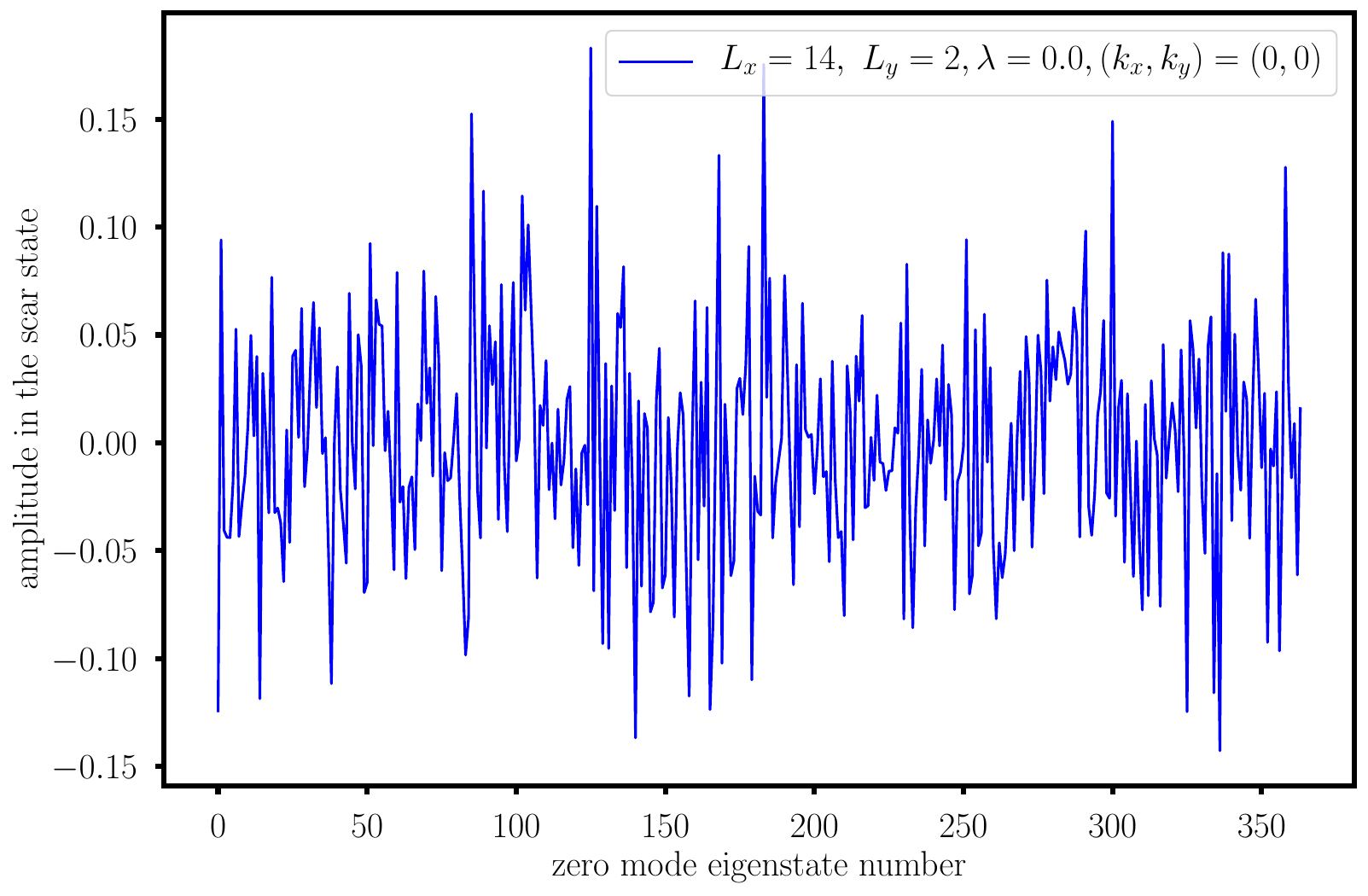}\\
   \includegraphics[width=0.43\textwidth]{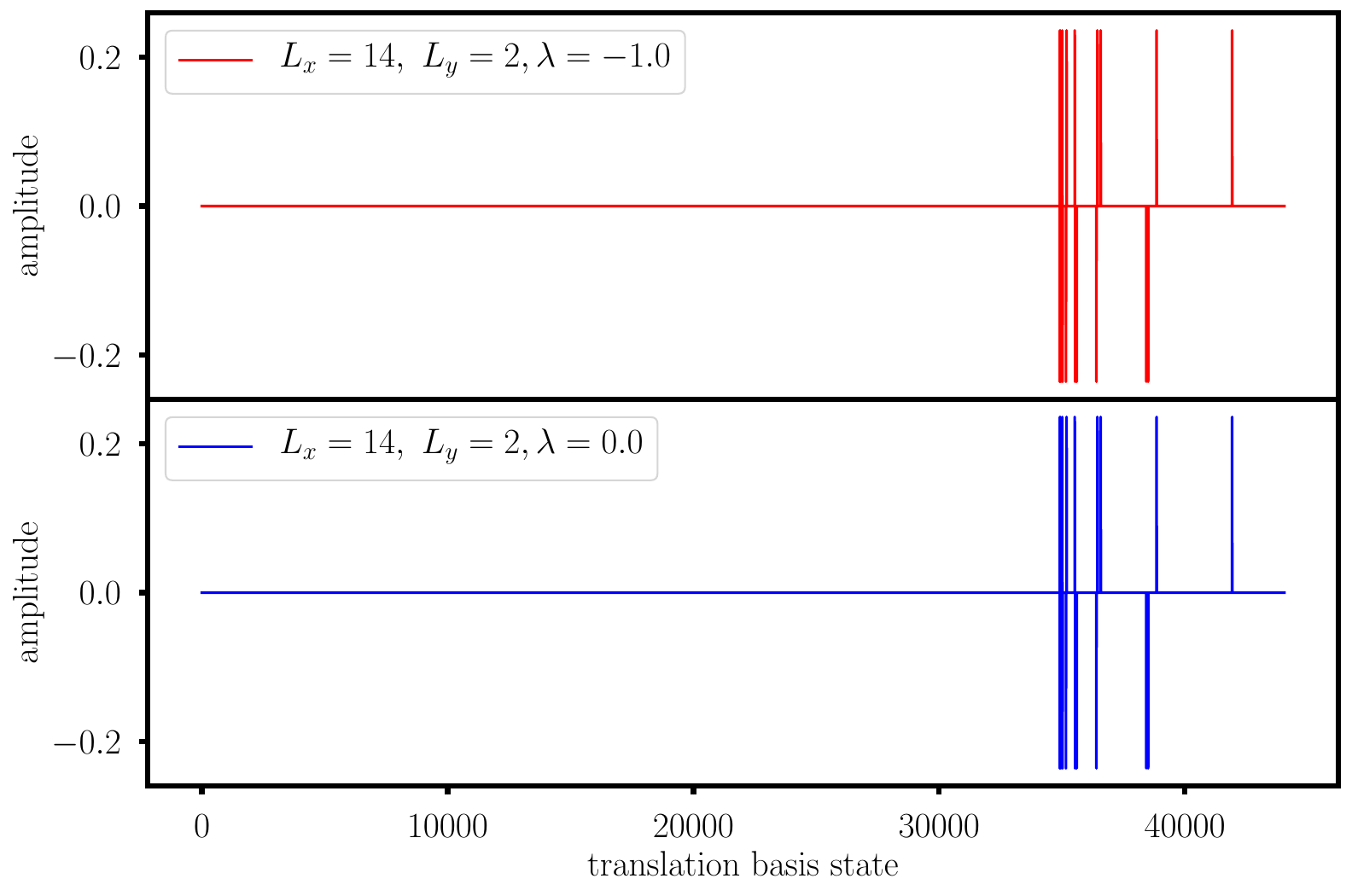}
   \caption{(Top panel) The QMBS at momentum $(k_x,k_y)=(0,0)$ expressed in
     terms of the zero modes for a ladder of dimension $(L_x,L_y)=(14,2)$.
     The amplitude of the QMBS expressed in the basis states of
     momentum $(k_x,k_y)=(0,0)$ where the the middle plot is directly obtained
     from the wavefunction of the QMBS at $\lambda=-1$ while the bottom plot
     is obtained from the eigenvector with integer eigenvalue of $\Opot$ in
     the subspace of the zero modes at $\lambda=0$.}
  \label{zeromodes2}
 \end{figure}

 Unlike the case of $L_y=2$ where the ladders have $4$ QMBS, one each at momenta 
 $(k_x,k_y)=(0,0),(\pi,\pi),(\pi,0), (0,\pi)$, the wider ladder of width $L_y=4$ have multiple scars at each of these four momenta ($10$ for $(k_x,k_y)=(0,0)$
 and $(\pi,\pi)$ respectively and $9$ for $(k_x,k_y)=(\pi,0)$ and $(0,\pi)$
 respectively). This is seen clearly from the behavior of the 
 Shannon entropy $S_1$ (Fig.~\ref{wideladder}, top left panel) and the electric flux correlator 
 (Fig.~\ref{wideladder}, bottom panel) for a ladder of dimension
 $(L_x,L_y)=(8,4)$ at $\lambda=-1$
 where the data shows outliers at $E=\lambda N_p/2$. As shown in Fig.~\ref{wideladder} (top right panel) for $(k_x,k_y)=(0,0)$, these $10$ scars are
 degenerate with energy $\lambda N_p/2$ and have a lower value of $S_1$ compared to that of other neighboring eigenstates with $E \neq \lambda N_p/2$. 

\section{QMBS from zero modes}

The QMBS at momentum $(k_x,k_y)=(0,0)$ obtained at $\lambda \neq 0$ for the ladder of dimension 
$(L_x,L_y)=(14,2)$ is a very particular linear combination of the mid-spectrum zero modes at 
$\lambda=0$ that also diagonalizes $\Opot$. However, from Fig.~\ref{zeromodes2} (top panel), 
the QMBS appears to be a pseudo-random superposition of all the zero modes which is, remarkably, 
stabilized at any finite $\lambda$. We have also verified that the matrix for $\Opot$, when 
expressed in the basis of the zero modes at momentum $(k_x,k_y)=(0,0)$, has a single eigenvalue 
which is an integer (that equals $N_p/2$). This wavefunction (Fig.~\ref{zeromodes2} (bottom panel)) 
is identical to the one obtained directly from the ED data for the $(k_x,k_y)=(0,0)$ QMBS
at $\lambda=-1$ (Fig.~\ref{zeromodes2} (middle panel)). Only $18$ out of the $44046$ basis states have 
non-zero coefficients while the other coefficients are (essentially) zero within numerical resolution. The number of basis states that contribute to the QMBS at momenta $(0,0), (\pi,\pi), (\pi,0),(0,\pi)$ for $(L_x,2)$ with $8 \leq L_x \leq 14$ is given below. The table immediately shows that these eigenstates are localized in the Hilbert space and have a low value of Shannon entropy $S_1$.

\begin{center}
  \begin{tabular}{|c|c|c|c|c|}
	  \hline
	  $L_x$ & Basis states & Basis states  &  Basis states & Basis states \\
          &in $(0,0)/(\pi,\pi)$ & in QMBS & in $(\pi,0)/(0,\pi)$ & in QMBS \\ 
	  \hline
          8  &     142 &      5     &       141 & 4   \\
          10 & 902 & 6 & 891 & 7 \\
          12 & 6166 & 13 & 6163 & 12 \\
          14 & 44046 & 18 & 43989 & 19 \\
	  \hline
  \end{tabular}
\end{center}

\end{document}